\documentstyle[twoside,fleqn,nuclphys,epsf]{article}

\newcommand{\nll}{\nonumber \\}

\newcommand{\bq}{\begin{equation}}
\newcommand{\eq}{\end{equation}}
\newcommand{\ba}{\begin{eqnarray}}
\newcommand{\ea}{\end{eqnarray}}
\newcommand{\req}[1]{(\ref{#1})}

\newcommand{\nobody}{\rule{0ex}{1ex}}

\newcommand{\AmS}{{\protect\the\textfont2
  A\kern-.1667em\lower.5ex\hbox{M}\kern-.125emS}}

\title{{\nobody\vspace{-23mm}\\ LMU 06/96}\hfill\vspace{10mm}\\
Four fermion final states in $e^+e^-$ annihilation}
\author{A. Leike
\address{Ludwig-Maximilians-Universit\"at,
Theresienstr. 37, D--80333 M\"unchen, Germany}
\thanks{Partially supported by EC contract CHRX-CT94-0579}
\thanks{Invited minireview held at ``QED and QCD at higher orders'', 
       Rheinsberg, April 1996.}
}
\begin{document}

\begin{abstract}
The present status of calculations of four fermion final states is reviewed. 
Higher order problems arising there are pointed out. 
Special attention is paid to results obtained in the semi-analytical
approach and their limitations and perspectives.
\end{abstract}

\maketitle

\section{Introduction}
Four fermion final states are already observed at LEP1 \cite{lep14f}.
They are the main final states to be investigated at LEP\,2 \cite{lep2}.
The two LEP runs planned this year 
\ba
\mbox{June/July\ \ at\ \ }  \sqrt{s}=161\,GeV \nll
\mbox{September/October\ \ at\ \ }  \sqrt{s}=174\,GeV \nonumber
\ea
will give the first opportunity to compare theoretical predictions
with the experimental data at the highest energies ever reached in $e^+e^-$
collisions.

The production of {\it two fermions} in the final state is quite simple at
the Born level. 
It proceeds through annihilation of gauge bosons in the $s$ channel,
figure~1a. 
For Bhabha scattering, the gauge boson exchange is also possible in
the $t$ channel, figure~1b. 

\begin{figure}[tbh]
\begin{center}
\hspace{-4cm}
\begin{minipage}[t]{7.8cm} {
\begin{center}
\ \vspace*{2.0cm}\\
\hspace{-1.7cm} \mbox{ \epsfysize=7.0cm \epsffile[0 0 300 300]{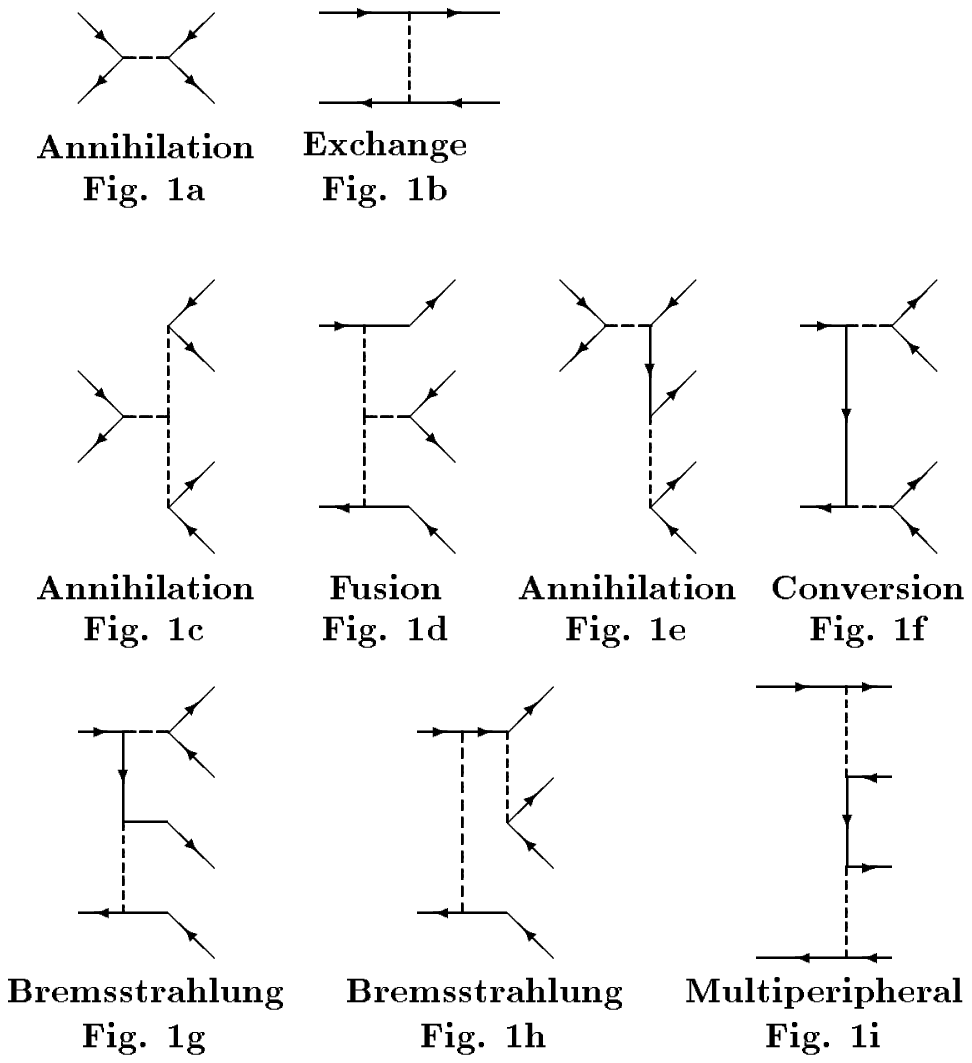}
}
\ \vspace*{-2.0cm}\\
\end{center}
}\end{minipage}
\end{center}
\noindent {\small\bf Fig.~1: }{\small\it 
Topologies of the lowest order Feynman diagrams for $e^+e^-\rightarrow
f\bar f$ and $e^+e^-\rightarrow f_1 f_2 f_3 f_4$. 
The dashed lines represent any allowed gauge boson.}
\end{figure}
%

{\it Four fermion} final states can be created only in higher order processes. 
They are described by sets of Feynman diagrams, which have a much
richer topology.
Two diagrams, figure~1c and 1d are essentially new containing the
non-abelian interaction of gauge bosons. 
The diagrams~1e and 1f can be interpreted as the
radiation of a fermion pair from the initial and final state of the
diagram~1a. 
Similarly figures~1g and 1h show pair radiation from the diagram~1b. 
The corresponding diagrams with Higgs exchange are not shown in
figure~1. 
They can be obtained by replacements of a gauge boson with the Higgs
at the appropriate places.

The cross sections of four fermion final states can be  
of the order of those of fermion pair production due to large factors,
which can arise and compensate the additional $\alpha^2$.

The diagrams in figures~1c and 1f contain two resonating 
gauge bosons, which tend to be on mass-shell. 
They are called signal diagrams for $W$ and $Z$ pair
production. Any resonating gauge boson $V$ enhances the cross section
by roughly a factor 
\bq
M_V/\Gamma_V\approx 37 \mbox{\ \ for\ }V=W \mbox{\ or\ }Z.
\eq
If the resonating boson is a photon going to $f\bar f$, the
enhancement factor (without cuts) is $\ln[s/(m(f)+m(\bar f))^2]$.
If the photon couples to light fermions, we recover the infrared
divergence due to soft photons. 
We assume that both fermions $f,\bar f$ are seen in the detector. 
This implies that the virtual photon cannot be arbitrarly soft.
The diagrams shown in figures~1d, 1e, 1g and 1h are enhanced by only
one resonating gauge boson.

Diagram~1e is unique because it is the only one, which allows gluon
exchange at the tree level. This gives an enhancement by a factor
\bq
\alpha_s^2(q^2)/\alpha^2\approx 14^2 \mbox{\ to\ }25^2 
\eq
depending on the scale $q^2$ of $\alpha_s$.

All diagrams with gauge boson exchange in the $t$ channel (figures~1b,
1d, 1g, 1h and 1i) contain collinear singularities.
Although they are regularized by the non-zero electron mass, 
an enhancement factor 
\bq
\ln s/m_e^2\approx 28
\eq
remains for every collinear photon in the total cross section.
This large factor can be reduced by a cut on the
angle between the outgoing $e^+(e^-)$ and the beam axis,

Let us be more explicit and estimate the total cross sections for 
$e^+e^-\rightarrow W^+W^-$ and $e^+e^-\rightarrow e^+e^-f\bar f$
compared to $e^+e^-\rightarrow f\bar f$ by collecting only
large enhancement factors.

$W$ pair production has a double resonant enhancement.
A further enhancement comes from the counting of the possible final
states. 
Fermion {\it pair production} results in 5 different quark flavours with
three different colors each or in 3 different lepton or neutrino pairs.
This adds up to 21 different final states. 
A $W$ can decay in 2 different quark pairs of 3 different colors each or in
3 different lepton pairs. This gives 9 different final states. We have
yet to take into account the combinatorical factor $\sqrt{2}$, which
enhances the $W^\pm f_1f_2$ couplings relative to the $Zf\bar f$
couplings. This is also the reason, why $\Gamma_Z/\Gamma_W\approx
21/(2\cdot 9)$ and not $21/9$.
Combined, we get the following additional factor for $W$ pair production
compared to fermion pair production:
\ba
\label{wwest}
&&\ \hspace{-1cm}
\sigma(e^+e^-\rightarrow (WW)\rightarrow f_1 f_2 f_3 f_4)
\nll
&\approx &\sum_f\sigma(e^+e^-\rightarrow f\bar f)\cdot 
\alpha^2\cdot \frac{(2\cdot 9)^2}{21}\frac{M_W^2}{\Gamma_W^2}\nll
&\approx &\sum_f\sigma(e^+e^-\rightarrow f\bar f)\cdot 1.4
\ea
After dividing out the threshold factor $\beta =\sqrt{1-4M_W^2/s}$ for
$W$ pair production, this crude estimate is correct within a factor 1.5
for $2M_W < \sqrt{s} < 500\,GeV$. 
For very high energies, the cross section for $W$ pair production 
becomes larger than the estimate \req{wwest} due to terms proportional to 
$\ln (s/M_W^2)$.
The production of two intermediate
$Z$ bosons is suppressed by the effective symmetrization
arising compared to $W$ pair production.

The cross section of the reaction $e^+e^-\rightarrow e^+e^-f\bar f$ 
is dominated by diagrams shown in figure~1i.
All three propagators in the $t$ channel can become singular
simultaneously leading to huge enhancement factors.
These factors can be estimated starting from the description of the process as
a scattering of photons with Weizs\"acker-Williams spectrum \cite{bm}.
Keeping only the leading terms, one gets
\ba
&&\ \hspace{-1cm}
\sigma(e^+e^-\rightarrow e^+e^-f\bar f)\\
&\approx &\sigma^{QED}(e^+e^-\rightarrow f\bar f)\nll
&&\cdot \frac{3\alpha^2Q_f^2}{4\pi^2}\frac{s}{m_f^2}
\ln\left(\frac{s}{m_e^2}\frac{s}{4m_f^2}\right)\ln\frac{s}{4m_f^2}
\ln\frac{4m_f^2}{m_e^2}.\nonumber
\ea
For $f=b$, we get at $\sqrt{s}=200\,GeV$
\ba
&&\ \hspace{-1cm}
\sigma(e^+e^-\rightarrow e^+e^-b\bar b)\nll
&\approx &\sigma^{QED}(e^+e^-\rightarrow b\bar b)\cdot 3\approx 2.5\,pb.
\ea
This is less than a factor 2 off from the exact result, see
figure~13 in \cite{bm}.

All four fermion final states can be classified according 
to the topologies of Feynman diagrams entering the process. 
The number of the Feynman diagrams involved in different processes with
charged current exchange is shown in table~1, while table~2 contains
final states, which are produced by the exchange of neutral gauge
bosons only. Potential diagrams with Higgs exchange are not counted. 
Diagrams coming from different contributions of quark mixing are counted.
Final states indicated in {\bf bold} contain only the topologies of
figures~1e and 1f. In the case of charged currents,
the topology~1c  has to be added. 
These charged current  processes 
are usually referred as CC9, CC10 or CC11 processes.
Numbers in {\it italics} indicate final states, which belong to table~1 and
table~2. The CC43 and CC19 (identical to NC43 and NC19) processes 
are described only by the topologies of the {\bf bold} processes.
However, here new structures in the interferences appear.
The CC56 (NC56) process involve all topologies. 
All remaining processes contain electrons
or electron neutrinos in the final state allowing gauge boson
exchange in the $t$ channel. The corresponding numbers are written in
roman. They contain all topologies simultaneously, except diagram~1d,
which demands at least one electron neutrino in the final state.
Final states with  identical final fermions are given in {\tt typewriter}.
They require additional contributions to insure Pauli antisymmetrization. 
\vspace{-1cm}{\small
\begin{table}[ht]
\begin{center}
\begin{tabular}{|c|c|c|c|c|c|}
\hline
             &
\raisebox{0.pt}[2.5ex][0.0ex]{${\bar d} u$}
& ${\bar s} c$ & ${\bar e} \nu_{e}$ &
              ${\bar \mu} \nu_{\mu}$ & ${\bar \tau} \nu_{\tau}$   \\
\hline
$d {\bar u}$            &{\it  43}& {\bf 11} &  20 & {\bf 10} & {\bf 10} \\
\hline
$e {\bar \nu}_{e}$      &  20 &  20 &{\it 56}&  18 &  18 \\
\hline
 $\mu {\bar \nu}_{\mu}$ & {\bf 10} & {\bf 10} &  18 & {\it 19} & {\bf 9}  \\
\hline
\end{tabular}
\end{center}
{\bf Tab.~1:} {\it Number of Feynman diagrams for} CC {\it type final
states, see \cite{gentle_unicc11}.}
\end{table}
\vspace{-1.5cm}
%
\begin{table}[ht]
\begin{center}
\begin{tabular}{|c|c|c|c|c|c|c|}
\hline
&
\raisebox{0.pt}[2.5ex][0.0ex]{${\bar d} d$}
&${\bar u} u$
&${\bar e} e$
&${\bar \mu} \mu$
&${\bar \nu}_{e} \nu_{e}$
&${\bar \nu}_{\mu} \nu_{\mu}$
\\
\hline
\raisebox{0.pt}[2.5ex][0.0ex]{${\bar d} d$}
 & {\tt 4$\cdot $16} & {\it 43} & {48}
             & {\bf 24} & 21 & {\bf 10} \\
\hline
\raisebox{0.pt}[2.5ex][0.0ex]
{${\bar s} s, {\bar b} b$} & {\bf 32} & {\it 43} & {48}
             & {\bf 24} & {21} & {\bf 10} \\
\hline
${\bar u} u$ & {\it 43} & {\tt 4$\cdot$16} & {48}
             & {\bf 24} & {21} & {\bf 10} \\
\hline
${\bar e} e$ &{48} &{48} & \textsf{4$\cdot$36} &{48}
& {\it 56} & {20}
\\
\hline
${\bar \mu} \mu$  & {\bf 24} & {\bf 24} & {48} & {\tt 4$\cdot$12}
                  & {19} & {\it 19}         \\
\hline
${\bar \tau} \tau$& {\bf 24} & {\bf 24} & {48} & {\bf 24}
                  & {19} & {\bf 10}         \\
\hline
${\bar \nu}_e \nu_{e}$  & {21} & {21} & {\it 56} & {19}
                  & \textsf{4$\cdot$9} & {12}                   \\
\hline
${\bar \nu}_{\mu} \nu_{\mu}$ & {\bf 10} & {\bf 10} & {20}
             & {\it 19} & {12} & {\tt 4$\cdot$3}  \\
\hline
${\bar \nu}_{\tau} \nu_{\tau}$ & {\bf 10} & {\bf 10} & {20}
             & {\bf 10} & {12} & {\bf 6}  \\
\hline
\end{tabular}
\end{center}
{\bf Tab.~2:} {\it Number of Feynman diagrams for} NC {\it type final
states, see \cite{gentle_nc24}.}
\end{table}
}\vspace{-1cm}

\section{Higher order corrections}
Four fermion final states can be produced with cross sections of
several $pb$.
With an integrated luminosity of $500pb^{-1}$ per year, 
the expected experimental precision for some cross sections is about 1\%. 
It has to be met by theoretical predictions, which have errors of 0.5\%
or less. Hence, higher order corrections have to be included. 

\subsection{QED corrections}
The complete QED corrections to four fermion final states are not known.
One could simplify the problem for $W$ pair production considering
corrections to the signal diagrams figures~1c and 1f only.
As an important step, this was done in the case of on-shell $W$'s \cite{jeger}.
For off-shell $W$'s, the complete correction is not known. 

In the following, the Coulomb singularity and
initial state corrections are mentioned.
\vspace{2mm}\hfill\\
{\bf Coulomb singularity}\hfill\\
The Coulomb singularity \cite{coulomb} arises from long range electromagnetic
interactions between the produced massive charged particles.
For $W$ pair production, we get the correction
\bq
\label{coul}
\sigma^{Coul} =\sigma^0\left(1+\frac{\alpha\pi}{2\beta}\right)
\eq
to the Born cross section $\sigma^0$, which diverges near threshold
where the velocity of the $W$'s $\beta=\sqrt{1-4M_W^2/s}$ approaches
zero. It indicates that perturbation theory is not applicable in this
region. 
Fortunately, the non-zero width $\Gamma_W$ and a slight off-shell production
of the $W$'s regularize the singularities in equation
(\ref{coul}).
The numerical effect can exceed 6\% of the total cross
section near threshold  \cite{fadin,cc11}. Therefore, the Coulomb
correction has to be inlcuded.

Formula (\ref{coul}) also applies for QED and QCD corrections to pair
production of massive fermions as it appears in the diagrams 1d-1h. 
This is well known from final state corrections to the
diagram~1a at LEP\,1 \cite{jlzabl}.
The problems can be cured by a calculation in the limit of massless
fermions or by a cut on the invariant mass of the heavy fermion pairs. 
Such a cut is desirable for quark pairs in any case to avoid
non-perturbative bound state regions.
\vspace{2mm}\hfill\\
{\bf Initial state corrections}\hfill\\
Initial state QED corrections to off-shell $W$ pair production are
calculated in \cite{gentle_nunicc}. 
They reach several \% near the $WW$ threshold.
In such calculations a problem arises because it is not clear 
how to separate weak corrections from QED corrections and 
initial state QED corrections in a unique and gauge invariant way. 
This separation problem is solved by the current splitting technique 
\cite{gentle_nunicc}, in which the
chargeless neutrino exchanged in the $t$ channel is divided into two
charge flows of opposite sign. Now the charge flows of the initial
and final states are separated, and gauge invariance is ensured as it
is in the case of $Z$ pair production.
The resulting initial state QED corrections can be split into universal
contributions, which are described by the same flux function as known
from annihilation diagrams in the case 
 of LEP\,1 physics, and into non-universal contributions 
depending on the particular process.
The non-universal contributions to off-shell $W$ and $Z$ production 
are numerically small for LEP\,2 energies. 
They are suppressed by a factor $s_1s_2/s^2$, where $s_{1,2}$ are 
the invariant energy flows through the $W$'s ($Z$'s)
\cite{gentle_nunicc,gentle_nuninc}. The smallness of the non-universal
corrections  cannot be taken for granted, but has to be proven
for any particular process.

The universal corrections lead to handy formulae for initial state
corrections to cross sections of $W$ and $Z$ pair production,
\bq
\sigma^{ISR}(s) = \int_{s'^-}^s
                  \frac{{\rm d}s'}{s}\sigma^0(s')\rho(s'/s).
\eq
Alternatively, initial state radiation can be taken into account by the
structure function approach \cite{kura}.
It assumes that both colliding photons have energies degraded by
radiated collinear photons,
\ba
&&\ \hspace{-1cm}
\sigma^{ISR}(s) \\  
&=&\hspace{-0.1cm}\int_{x_1^-}^1{\rm d}x_1\int_{x_2^-}^1{\rm d}x_2
                  D(x_1,s)D(x_2,s)\sigma^0(x_1x_2s).\nonumber
\ea
The flux function (FF) $\rho(s'/s)$ and the structure function (SF)
$D(x,s)$
contain information about real and virtual corrections.
Further details can be found, for example, in \cite{cc11} or \cite{bkp}.

\subsection{Electroweak corrections}
Complete one-loop corrections to four fermion processes would be
the best task to get more accurate descriptions of these cross
sections.
Unfortunately, this calculation is very complex \cite{old} and done
only for on-shell $W$'s \cite{jeger}.

In this paragraph, only electroweak corrections connected with the
finite widths of gauge bosons are mentioned.

At lowest order, a particle $V$ has no width, i.e. its propagator $D_V$ is
proportional to $D_V\sim [q^2-M_V^2]^{-1}$, where $q^2$ is the momentum
transfer through the particle of mass $M_V$. 

Massive gauge bosons are unstable having a non-negligible decay width. 
The required precision at the energy range of LEP\,2 
demands the inclusion of a finite width.
Although the inclusion of a width for particles in the $t$ channel is
not physical,
the simplest possibility is the inclusion of the constant on-shell
decay width $\Gamma_V$ everywhere in the propagator, 
$D_V\sim [q^2-M_V^2+iM_V\Gamma_V]^{-1}$.

The next step of accuracy, which cures this deficiency, 
is to take into account the natural energy dependence
of the width arising from self-energy insertions into the boson propagator.
Present and future experiments are sensitive to the difference of
these handlings of the width \cite{ztrafo,bd}, even at the limit of
very high energies.

A finite width in the propagator implies the partial 
inclusion of contributions, which are of higher order in perturbation
theory. 
Unfortunately, the symmetries of a theory (as gauge invariance) 
are only respected in a fixed order of perturbation theory.
Hence, the inclusion of a non-zero width leads to gauge dependent
cross sections.
Although the gauge dependent terms are of higher order in perturbation
theory, they can be enhanced by large kinematical factors as $s/m_e^2$
in the reaction $e^+e^-\rightarrow e^-\bar\nu_e u\bar d$, see
\cite{gaugeviol1}.
Therefore, finite widths must be included with care. 
Different approaches for the restoration of gauge invariance 
have been proposed, among them the Laurent expansion of the matrix
element and the inclusion of projection operators \cite{stuart} or
the inclusion of certain higher order contributions \cite{gaugeviol2}. 

\subsection{QCD corrections}
QCD corrections give sizeable contributions to distributions and cross
sections. 
They are known from LEP\,1 for the production
of {\it pairs} of (heavy) quarks \cite{jlzabl}.
They are under investigation for {\it four quarks} in the final state
\cite{pittau}. 

QCD corrections enter the width of the $Z$, $W$ and the Higgs.
For the case of $\Gamma_W$, QCD corrections can reach several \%, see 
\cite{bakl}.
Most of the QCD corrections to the Higgs width can be absorbed into
the running quark masses evaluated at the scale of the Higgs mass.
They lead to values for $m_b(M_H)=2.9\,GeV$ and $m_c(M_H)=0.6\,GeV$
\cite{higgs}, which deviate substantially from the corresponding pole masses. 

\subsection{Theoretical uncertainties}
The theoretical uncertainties are extensively discussed in the
different contributions to \cite{lep2}.
Here, we only give some of them in a telegraphic style.\vspace{2mm}\hfill\\
 {\bf QED corrections}:\hfill\\
$\bullet$ 
The initial state QED corrections to {\it all} Born cross sections presented
in \cite{lep2} are calculated by the FF or by the SF approach. 
The validity of these approximations is proven only
for $W$ and $Z$ pair production and for annihilation diagrams. 
For other processes it is not known, whether
the non-universal corrections are small enough to be neglected.
Even for $W$ and $Z$ pair production, they reach 0.3\% at LEP\,2
energies and rise above 1\% at 1\,TeV
\cite{gentle_nunicc,gentle_nuninc}.\hfill\\
$\bullet$ 
The FF is derived after an average over all phase
space variables except the two invariant energy flows $s_{1,2}$
through the $W$'s or $Z$'s. For distributions or cuts in parameters
different from $s_1$ and $s_2$ (i.e. angular cuts, cuts on energies of
single particles), it depends in general
on the additional parameters. This is known from LEP\,1, where the
angular-dependent FF for annihilation diagrams is derived \cite{666}.
The dependence on the distribution parameter disappears only for soft 
photon radiation.
Therefore, the theoretical errors of observables, which favor 
hard photon radiation can become very large. 
Fortunately, these observables have small cross sections because
QED favors soft photon radiation.
\hfill\\
$\bullet$ 
We now assume that the FF and SF approaches work
perfectly to all orders of perturbation theory. 
We can then use our knowledge about these functions beyond the leading
order and study the changes of cross sections.
This relative change reaches 0.5\% \cite{bakl} giving an idea about the
effect of higher order QED corrections.\vspace{2mm}\hfill\\
{\bf QCD corrections:}\hfill\\
$\bullet$
A calculation of complete QCD corrections to four fermion processes 
is not done. 
The resulting uncertainty can be estimated in the following simple example:
$W$ pair production with four quarks in the final state involves 
diagram~1e with gluon exchange at the Born level. 
The diagram favors soft gluons. The minimal momentum transfer through
the gluon is usually defined by cuts. 
The scale $q$ of $\alpha_s(q^2)$ in these contributions is not known exactly.
An uncertainty of 20\% in the
scale $q$ leads to an uncertainty of 10\% in $\alpha_s^2(q^2)$ for $q\approx
10\,GeV$. If the diagram~1e would contribute 10\% to the total cross
section, this scale uncertainty would transform to an error of 1\% in the
total cross section. If the cuts allow smaller $q^2$, 
$\alpha_s(q^2)$ and its uncertainty become larger. This simple example
illustrates the dependence of QCD corrections on kinematical cuts.
\hfill\\
$\bullet$ 
The running quark masses in Higgs decay depend on $\alpha_s(M_Z)$. The
resulting uncertainties are shown in table~3 in \cite{higgs}. They
reach several \%. They transform into uncertainties of
predictions for production cross sections of Higgs Bosons being
proportional to the square of these masses.
\hfill\\
$\bullet$
The interface between partonic cross sections and
hadronization procedures and the hadronization procedures itself
introduce  theoretical uncertainties.\vspace{2mm}\hfill\\
{\bf Weak corrections}\hfill\\
Gauge invariance can be restored in calculations with finite widths by
different methods. The arising numerical difference is studied for the
process $e^-e^+\rightarrow e^-\bar\nu_eu\bar d$ in the second
reference of \cite{gaugeviol2} and found to be about 0.5\%. 
The disagreement between the different methods is
expected to be larger near threshold \cite{old}.\vspace{2mm}\hfill\\
{\bf Final remark}\hfill\\
The estimation of theoretical errors is a highly subjective task. 
These errors can be reduced only by future higher order calculations.
Remembering that not all
theoretical errors are equally important in all observables, we 
conclude that the present theoretical accuracy probably meets the
precision expected in the first year at LEP\,2. 
It has certainly to be improved in the future.

\section{Codes and algorithms}
\subsection{Existing codes}
Event generators for four fermion final states are described in detail
in volume 2 of reference \cite{lep2}. They use different algorithms
for the evaluation of the squared matrix elements, phase space
integration and mapping of singularities. 
Numerical comparisons between the programs gave a nice agreement.

It follows a list of the available codes and its authors.\vspace{2mm}\hfill\\
{\bf Monte Carlo Programs:}\hfill\\
$\bullet$ ALPHA (F. Caravaglios, M. Moretti)\hfill\\
$\bullet$ CompHep 3.0 (E. Boos, et al.)\hfill\\
$\bullet$ ERATO (C.G. Papadopoulos)\hfill\\
$\bullet$ EXCALIBUR (F.A. Berends, R. Kleiss,\hfill\\ 
\nobody\hspace{3mm}R. Pittau)\hfill\\
$\bullet$ grc4f 1.0 (J. Fujimoto, et al.)\hfill\\
$\bullet$ KORALW 1.03 (M. Skrzypek, S. Jadach, \hfill\\
\nobody\hspace{3mm}W. P{\l}aczek, Z. W\c{a}s) \hfill\\
$\bullet$ LEPWW (F.C. Ern\'e)\hfill\\
$\bullet$ LPWW02 (R. Miquel, M. Schmitt)\hfill\\
$\bullet$ PYTHIA 5.719 / JETSET 7.4 (T. Sj\"ostrand)\hfill\\
$\bullet$ WOPPER1.4 (H. Anlauf, T. Ohl)\hfill\\
$\bullet$ WPHACT (E. Accomando, A. Ballestrero)\hfill\\
$\bullet$ WWF 2.2 (G.F. van Oldenborgh)\hfill\\
$\bullet$ WWGENPV/HIGGSPV (G. Montagna, \hfill\\
\nobody\hspace{3mm}O. Nicrosini, F. Piccinini)
\vspace{2mm}\hfill\\
{\bf Other codes:}\hfill\\
$\bullet$ GENTLE/4fan (D. Bardin, D. Lehner, A. Leike,\hfill\\
\nobody\hspace{3mm}T. Riemann),
Semi-analytical code\hfill\\
$\bullet$ WTO (G. Passarino), multi-dimensional \hfill\\
\nobody\hspace{3mm}deterministic integration

\subsection{Comparison between the Monte Carlo the and semi-analytical
approach}
Many different topologies of Feynman diagrams are involved in
calculations of four fermion processes leading to a rather
singular matrix element.
It has to be integrated over the 8-dimensional phase space.
In the case of absence of transversal beam polarization,
one degree of freedom, the rotation around the beam axis, is
trivial. We are left with a 7-dimensional phase space integration. 
In the semi-analytical (SA) approach, most of the phase space
integrations are done analytically. The remaining integrals are calculated
numerically. In the Monte Carlo (MC) approach, all integrals are taken
numerically. 

As shown in section~2, kinematical cuts are needed to reject threshold
regions of massive fermion pairs because they are not described by 
perturbation theory. 
Additional angular and energy cuts are required to ensure the
detection of four final particles in a real detector. 
Finally, a real detector has wholes, which are planned 
(i.e. for cables and support) and those which arise spontaneously 
(i.e. damaged and dead sectors). 
These individual properties have to be inlcuded into a data analysis. 
There is an additional physical motivation for cuts to separate
interesting events from the background. 

Having in mind these requirements, it is obvious that one needs a MC
 event generator. It is most flexible in kinematical cuts and can
be connected with detector simulations and hadronization procedures.

On the other hand, fits require fast and accurate programs. 
The overall accuracy (which consists of theoretical uncertainties and 
numerical uncertainties)
must not be larger than half of the experimental error. 
Remembering the present theoretical errors, it follows that the numerical 
uncertainty should be considerably smaller than the experimental error.
Demanding that a fit to future data should be finished rather in 1 to
10 days and not in 10 to 100 days, and assuming 100 to 1000 cross
section calculations for one fit, one can estimate the 
calculation speed and accuracy of a code, which is useful for fits:
Such a code should calculate one cross section with an numerical accuracy
of $\le 0.1$\% in less than 15 minutes on computers already available today. 
The higher speed of future computers will probably be eaten by the
implementation of future theoretical results needed to decrease 
the theoretical errors.

These requirements can be met by a SA code. In the simplest
case of the NC32 processes (all {\bf bold} final states in table~2),
the typical calculation time is 30s with a numerical accuracy of
0.1\% and with QED
corrections included by the FF approach. This time can vary
by a factor 3 depending on the cuts and on the final state. 
The calculation time for CC11 processes (all {\bf bold} final states in
table~1) is about 3 times longer due to the longer analytical
formulae involved. The codes of GENTLE/4fan distributed during the LEP\,2
workshop are much slower. This will be improved with an update at the
end of this year. 

SA codes are not as flexible as MC's allowing only the calculations of
a limited number of distributions. 
However, they establish a benchmark
for MC's because SA codes have an inherently much smaller numerical error. 
Unfortunately, the SA method can fail for certain processes.
For details, see the next section. 

The calculation time of SA and MC programs  depends on the numerical
accuracy $\varepsilon$ required. In the SA approach, we have an
additional dependence on the dimension $d$ of the numerical integration and
on the order $r$ of the numerical integration algorithm:
\ba
\label{calctime}
&&\mbox{\bf MC:\ \ } t\sim\varepsilon^{-2}\nll
&&\mbox{\bf SA:\ \ } t\sim\varepsilon^{-d/r}
\ea
A few remarks are in order inspecting relations (\ref{calctime}):\hfill\\
$\bullet$
In practice, the calculation time for MC's depends on $d$
because the integrand cannot be mapped to a completely flat function.\hfill\\ 
$\bullet$
High accuracy demands large calculation times in the MC approach. 
The difference between $\varepsilon =1\%$ and $\varepsilon =0.1\%$
has to be paid by a factor 100 in computer time.\hfill\\
$\bullet$
In the SA approach, we have $r=5$ for Simpson's rule. 
More sophisticated algorithms with larger $r$ don't
necessarily lead to shorter calculation times because their
errors are proportional to higher derivatives of the integrand.\hfill\\
$\bullet$
SA codes can easily achieve a high accuracy. They are very fast for
small $d$.
They fail for large $d$. In practice, the calculation time increases
by more than a factor 10 for every additional integration. 
We have $d=2,3,4$ or $5$ in GENTLE/4fan for calculations of 
Born cross sections, QED FF corrected cross sections, QED SF
corrected cross sections, QED SF corrected angular distributions.
A calculation time below 15 min seems to be possible for $d=3$ for 
all solvable processes, and for $d=4$ for selected processes only.

\section{The semi-analytical approach}
SA calculations result to nice analytical formulae, which depend only on
the input parameters and on the remaining integration variables. 
These formulae show symmetries, which allow a deeper physical 
understanding of the underlying process.

Specific four fermion processes are given by the
imaginary part of three-loop diagrams. 
Therefore, some formulae obtained in the SA approach agree with
results of three-loop calculations.
However, one has to admit flavour
identification in the four fermion final states. 
Therefore, one is not summing over complete weak multiplets.  
Furthermore, one is not interested in a {\it complete} analytical phase
space integration because one wants to calculate distributions and apply cuts.

In contrast to the MC, 
the SA approach can distinguish analytical zeros from contributions, which are
multiplied by (very) small factors. This found an interesting application
in the discussion of the interferences in SM Higgs production 
\cite{gentle_nc24h} and can be extended to SUSY Higgs production. 

\subsection{Cross sections, distributions and cuts}
So far, the SA approach 
\cite{gentle_nunicc,gentle_nc24,gentle_nc24h,gentle_unicc11} 
uses the following parametrization of the phase space
\footnote{Other parametrizations are considered for different topologies of
Feynman diagrams.}:
\ba
d\Omega
&=& \prod_{i=1}^4\frac{d^3p_i}{2p_{i}^0}
\delta^4(k_1+k_2-\sum_{i=1}^4 p_i)
\\
&=&\frac{\sqrt{\lambda(s,s_1,s_2)}}{8s}
\frac{\sqrt{\lambda(s_1,m_1^2,m_2^2)}}{8s_1}
\nll
&&\times\frac{\sqrt{\lambda(s_2,m_3^2,m_4^2)}}{8s_2}
d s_1 d s_2 d \Omega_0 d \Omega_1 d \Omega_2,
\nonumber
\end{eqnarray}
$k_1$ and $k_2$ are the four-momenta of the initial
electron and positron.
The fermions $f_i$ in the final state have four-momenta $p_i$ and masses $m_i$.
The invariants $s, s_1$, and $s_2$ are
\begin{eqnarray}
s&=&(k_1+k_2)^2,
\nll
s_1&=&(p_1+p_2)^2,
\hspace{0.5cm}
 s_2=(p_3+p_4)^2.
\end{eqnarray}
One has to substitute $ d \Omega_0=2\pi d\cos\theta_0=2\pi dc_0$ 
to integrate over the rotation angle around the beam axis.
$\theta_0$ is the angle between the vectors ($\vec{p}_1+\vec{p}_2$) and
$\vec{k}_1$.
The spherical angles of the momenta $\vec{p_1}$ and $\vec{p_2}$ 
($\vec{p_3}$ and $\vec{p_4}$)
in their rest frames are $\Omega_1\ (\Omega_2)$:
$d \Omega_i = d \cos\theta_i d \phi_i= dc_id\phi_i$.
The kinematical ranges of the integration variables are:
\ba
(m_1+m_2)^2  &\le& s_1    \le (\sqrt{s}-m_3-m_4)^2,
\nll
(m_3+m_4)^2 &\le& s_2 \le (\sqrt{s}-\sqrt{s_1})^2,
\nll
-1 &\le& c_i \le 1,
\nll
0 &\le& \phi_i\le 2\pi,\ \ \ i=0,1,2.
\ea

Usually, the squared matrix element is integrated over all six angles
$c_i,\phi_i$. The remaining two integrations are done numerically.
As a result, the cross section is obtained in the following form:
\ba
\label{sig}
\sigma(s)=\int_{s_2^-}^{s_2^+} d s_2 \int_{s_1^-}^{s_1^+} d s_1
\frac{d^2\sigma}{d s_1 d s_2}
\ea
The boundaries of the integrations $s_{1,2}^\pm$ allow the implementation of
cuts in $s_1$ and $s_2$. 
These cuts are very important to separate
intermediate photons, $Z$'s, $W$'s and Higgs bosons in the diagrams~1c
and 1f from the background. 
The double differential cross section factorizes,
\ba
\frac{d^2\sigma}{d s_1 d s_2}
&=&\sum_i C_i(e,f_1,f_2,f_3,f_4,s,s_1,s_2)\nll
&&\hspace{0.5cm}\times G_i(s,s_1,s_2),
\ea
where $C_i(\dots)$ are functions depending on the couplings and the
invariants and $G_i(\dots)$ are kinematical functions resulting from
the six-fold analytical integration. They depend on the invariants only. 
The summation runs over the interferences of different {\it topologies}.
The double differential cross sections allow the calculation of the
distributions 
\bq
\label{dists}
\frac{d\sigma}{ds_1},\ \ \ 
\frac{d\sigma}{ds_2},\ \ \ 
\frac{d\sigma}{dE_{12}},
\eq
where $E_{12}$ is the sum of the energies of the particles $f_1$ and  $f_2$,
\ba
E_{12}&=&p_{10}+p_{20}=\frac{s+s_1-s_2}{2\sqrt{s}}.
\ea
Cuts are possible on all parameters in equation (\ref{dists})
simultaneously. 

To be more differential, the integration over $c_0$ can be
left for numerical integration \cite{Le}. 
The results are triple differential cross sections,
\ba
\label{c01}
\frac{d^3\sigma}{d s_1 d s_2 d c_0}
&=&\sum_i C_i(e,f_1,f_2,f_3,f_4,s,s_1,s_2)\nll
&&\hspace{0.5cm}\times G_i(s,s_1,s_2,c_0)
\ea
with the remaining numerical integrations
\ba
\label{c02}
\sigma(s)=\int_{s_2^-}^{s_2^+} d s_2 \int_{s_1^-}^{s_1^+} d s_1
\int_{c_0^-}^{c_0^+} d c_0
\frac{d^3\sigma}{d s_1 d s_2 d c_0}.
\ea
The parameters $s_1,s_2$ and $c_0$ allow the construction 
of additional distributions, 
\bq
\label{dists0}
\frac{d\sigma}{dc_0},\ \ \ 
\frac{d\sigma}{dp_{12}^T},\ \ \ 
\frac{d\sigma}{dy_{12}},
\eq
where $p_{12}^T$ is the sum of the transversal momenta of the particles 
$f_1$ and $f_2$ against the beam axis, and $y_{12}$ is the rapidity
related to $p_{12}^T$.
The new parameters depend on $s_1,s_2$ and $c_0$ only,
\ba
p_{12}^T&=&\sqrt{E_{12}^2-(m_1+m_2)^2}\sqrt{1-c_0^2}\nll
\tanh y_{12}&=&c_0\sqrt{1-(m_1+m_2)^2/E_{12}^2}.
\ea
Again, cuts are possible on all parameters in equations (\ref{dists})
and (\ref{dists0}) simultaneously.

All distributions listed in equations (\ref{dists}) and (\ref{dists0}) 
can be calculated with initial state radiation too.
This is possible because $s_1$ and $s_2$ are invariants, and $c_0$ can 
be reconstructed after a boost due to photon radiation
from the initial state without use of $\Omega_1$ and $\Omega_2$.

Alternatively, one of the parameters $c_1$ or $c_2$ can be left for
numerical integration. The corresponding cross section formulae are
identical to (\ref{c01}) and (\ref{c02}) with $c_0$ substituted by
$c_1$ or $c_2$.
No new functions $C_i(\dots)$ and $G_i(\dots)$ appear in the case of 
NC32 processes.
The distributions
\bq
\label{dists1}
\frac{d\sigma}{dp_{i0}},\ \ 
\frac{d\sigma}{dp_i^T},\ \ 
\frac{d\sigma}{dy_i},\ \ 
\frac{d\sigma}{dc_{12}},\ \ \ i=1,2
\eq
are now calculable because
\ba
\label{borndists}
p_{10}&=&\frac{(s_1+m_1^2-m_2^2)(s+s_1-s_2)}{4s_1\sqrt{s}}\nll
   &&+
c_1\,\frac{\sqrt{\lambda(s_1,m_1^2,m_2^2)\lambda(s,s_1,s_2)}}
{4s_1\sqrt{s}},\nll
p_1^T&=&\sqrt{p_{10}^2-m_1^2}\sqrt{1-c_1^2}=p_2^T,\nll
\tanh y_1&=&c_1\sqrt{1-m_1^2/p_{10}^2}\mbox{\ \ and\ \ }\\
c_{12}&=&-\frac{1-(1-c_1^2)\lambda(s,s_1,s_2)/(4ss_1)}
               {1+(1-c_1^2)\lambda(s,s_1,s_2)/(4ss_1)}.\nonumber
\ea
depend on $s_1,s_2$ and $c_1$ only.
$p_1^T$ is the transversal momentum against the $\vec p_1+\vec p_2$
axis, $y_1$ is the rapidity related to $p_1^T$, 
and $c_{12}$ is the cosine of the angle between $\vec p_1$ and $\vec p_2$.
For simplicity, the formula for $c_{12}$ is given only in the massless limit.
As in the case with open $c_0$, cuts are possible to
the parameters in equations (\ref{dists})
and (\ref{dists1}) simultaneously.
Similar formulae can be written for the case where $c_2$ is left for
numerical integration.

The distributions (\ref{borndists}) can be calculated in the SA
approach only at the Born level. 
The reason is that $c_1 (c_2)$ cannot be reconstructed after a boost of
the subsystem due to photon radiation without knowledge of $c_0$ and 
$\phi_1 (\phi_2)$.
Unfortunately, these angles are already integrated out. 

\subsection{Limitations and perspectives}
Presently, only final states marked in {\bf bold} in tables~1 and 2
(NC32, CC11) are treated in the SA approach. 
Final states printed in the tables in roman (NC48, CC20) are
under investigation \cite{BaBiLeRi}. 
The NC43, NC19 (CC43, CC19) can certainly be treated.

All processes with identical particles in the final state (NC 4$\cdot$ 36) 
given in table~2 in {\tt typewriter} cannot be
treated with cuts on $s_1=(p_1+p_2)^2$ and $s_2=(p_3+p_4)^2$.
These cuts imply the same cuts on 
$\bar s_1=(p_1+p_4)^2$ and $\bar s_2=(p_3+p_2)^2$ because the
identical final particles are indistinguishable. 
They have to be calculated
with a phase space parametrization, which has the four
invariants $s_1,s_2,\bar s_1,\bar s_2$ as integration parameters. 
Unfortunately, the transformation to these parameters is described by general 
polynomials of fourth power, which have to be inverted. 

The four jet cross section with two quarks and two gluons in the final
state \cite{JaLe} has to be added incoherently to the NC32 and CC11 processes.
All soft singularities can be eliminated by a cut on the invariants
$s_1$ and $s_2$. The collinear singularities are regularized by a
finite quark mass $m_q$. However, the differential cross section is enhanced
by a factor $s/m_q^2$ for every gluon, which is collinear with a
quark. Experimentally, one can discover a four jet event only, if the
jets are separated by a minimal angle. Such an angular cut is not
possible in the SA approach. 
This shows another limitation of this calculation scheme. 
The problem arises already in four quark final final states but
without collinear enhancement. 

Finally, I would like to add some remarks about possible future SA
calculations. 

The inclusion of anomalous couplings is possible for the CC11 process
\cite{BiRi}. 
The treatment of SUSY Higgs cross section is possible everywhere,
where the corresponding topology with SM gauge bosons can be calculated.
This includes the $s$ channel Higgs diagrams in muon colliders.
Four fermion final states in $e^-e^-$ collisions
can be calculated, if the corresponding topologies are calculable in 
$e^+e^-$ collisions. $\gamma e (\gamma\gamma)$ collisions
demand one more integration for every photon in the initial state.
This complication could be compensated by the missing QED corrections
from the initial photon.
The inclusion of SUSY processes with four fermions in the final
states is interesting. 
Unfortunately, many SUSY processes lead to six fermion final states.

To summarize this section, we emphasize that the SA and MC approach
are complementary. 
Both approaches have advantages and disadvantages.
Both are needed at different places in a future data analysis.

\section{Summary}
The physics of four fermion final states is much richer than that of
two fermion final states.
The calculation of four fermion cross sections with a
theoretical error below 0.5\% is challenging.

In general, complicated kinematical cuts are required already for Born
cross sections to ensure the applicability of perturbation theory and 
to separate interesting events from the background. 
Radiative corrections are necessary.
The codes used in fits to future data must be fast and accurate.

Not all these features can be met by a code based on {\it one }
calculation scheme only.
Flexible Monte Carlo programs and fast semi-analytical programs together
could fulfill all requirements. 

Present codes probably meet the experimental precision expected at
LEP\,2 {\it this} year.
Theoretical errors certainly must be reduced to meet the accuracy of 
the LEP\,2 data in the future.


\end{document}